\documentclass[twocolumn]{aastex7}

\usepackage{CJK}
\usepackage{url}
\usepackage{hyperref}

\shorttitle{SMILES NIRSpec Spectra}
\shortauthors{Zhu et al.}

\begin{document}

\title{SMILES Data Release II: Probing Galaxy Evolution during Cosmic Noon and Beyond with NIRSpec Medium-Resolution Spectra}

\author[0000-0003-3307-7525]{Yongda Zhu}
\affiliation{Steward Observatory, University of Arizona, 933 North Cherry Avenue, Tucson, AZ 85721, USA}
\email[show]{yongdaz@arizona.edu}

\author[0000-0001-8470-7094]{Nina Bonaventura}
\affiliation{Steward Observatory, University of Arizona, 933 North Cherry Avenue, Tucson, AZ 85721, USA}
\email{}

\author[0000-0001-6561-9443]{Yang Sun}
\affiliation{Steward Observatory, University of Arizona, 933 North Cherry Avenue, Tucson, AZ 85721, USA}
\email{}

\author[0000-0003-2303-6519]{George H.\ Rieke}
\affiliation{Steward Observatory, University of Arizona, 933 North Cherry Avenue, Tucson, AZ 85721, USA}
\email{grieke@arizona.edu}

\author[0000-0002-8909-8782]{Stacey Alberts}
\affiliation{Steward Observatory, University of Arizona,
933 North Cherry Avenue, Tucson, AZ 85721, USA}
\affiliation{AURA for the European Space Agency (ESA), Space Telescope Science Institute, 3700 San Martin Dr., Baltimore, MD 21218, USA}
\email{salberts@stsci.edu}

\author[0000-0002-6221-1829]{Jianwei Lyu}
\affiliation{Steward Observatory, University of Arizona,
933 North Cherry Avenue, Tucson, AZ 85721, USA}
\email{}

\author[0000-0003-4702-7561]{Irene Shivaei} 
\affiliation{Centro de Astrobiolog\'ia (CAB), CSIC-INTA, Ctra. de Ajalvir km 4, Torrej\'on de Ardoz, E-28850, Madrid, Spain}
\email{}

\author[0000-0002-9288-9235]{Jane Morrison}
\affiliation{Steward Observatory, University of Arizona, 933 North Cherry Avenue, Tucson, AZ 85721, USA}
\email{}

\author[0000-0001-7673-2257]{Zhiyuan Ji}
\affiliation{Steward Observatory, University of Arizona, 933 North Cherry Avenue, Tucson, AZ 85721, USA}
\email{}

\author[0000-0003-1344-9475]{Eiichi Egami}
\affiliation{Steward Observatory, University of Arizona, 933 North Cherry Avenue, Tucson, AZ 85721, USA}
\email{}

\author[0000-0003-4337-6211]{Jakob M.\ Helton}
\affiliation{Steward Observatory, University of Arizona, 933 North Cherry Avenue, Tucson, AZ 85721, USA}
\email{}

\author[0000-0002-7893-6170]{Marcia J.\ Rieke}
\affiliation{Steward Observatory, University of Arizona, 933 North Cherry Avenue, Tucson, AZ 85721, USA}
\email{}

\author[0000-0002-5104-8245]{Pierluigi Rinaldi}
\affiliation{Steward Observatory, University of Arizona, 933 North Cherry Avenue, Tucson, AZ 85721, USA}
\email{}

\author[0000-0002-4622-6617]{Fengwu Sun}
\affiliation{Center for Astrophysics $|$ Harvard \& Smithsonian, 60 Garden St., Cambridge, MA 02138, USA}
\affiliation{Steward Observatory, University of Arizona, 933 North Cherry Avenue, Tucson, AZ 85721, USA}
\email{fengwu.sun@cfa.harvard.edu}

\author[0000-0001-9262-9997]{Christopher N.\ A.\ Willmer}
\affiliation{Steward Observatory, University of Arizona, 933 North Cherry Avenue, Tucson, AZ 85721, USA}
\email{}

\begin{abstract}
We present the second data release of the Systematic Mid-Infrared Instrument (MIRI) Legacy Extragalactic Survey (SMILES), focusing on JWST/NIRSpec medium-resolution spectroscopy of galaxies across cosmic time. This release includes spectroscopic observations of 166 galaxies spanning $0 < z < 7.5$, sampling star-forming galaxies, quiescent systems, and active galactic nuclei (AGN), with an emphasis on galaxies at cosmic noon ($z \sim 1$-3). We describe the target selection strategy, the observational setup with the G140M/F100LP and G235M/F170LP gratings, and the data calibration process. The final data products include the reduced spectra, redshift catalog, emission-line catalogs produced with \texttt{GELATO} for emission-line galaxies and \texttt{pPXF} fits for quiescent systems, and ancillary spectral energy distribution (SED) fit results derived from multi-band photometry. The SMILES NIRSpec dataset enables investigations of obscured AGN, multi-phase outflows, ionizing properties, and the role of environment in galaxy evolution. All data products are publicly available through STScI/MAST at \url{https://archive.stsci.edu/hlsp/smiles}.
\footnote{For peer review, the data products can be found at \url{https://github.com/ydzhuastro/smiles_dr2}}
\end{abstract}


\section{Introduction}

The period between redshift $z\sim1$ and 3, often referred to as ``cosmic noon,'' marks an important era in galaxy evolution. During this time, the universe experienced its peak star formation activity, rapid black hole growth, and the emergence of 
\added{ordered galaxy structures such as disks and bulges}
\citep{madau_cosmic_2014, forster_schreiber_star-forming_2020,espejo_salcedo_galaxy_2025,suess_rest-frame_2022,fudamoto_red_2022}. Investigating the physical conditions of galaxies during this epoch is essential to understand the build-up of stellar mass, the enrichment of the interstellar medium (ISM), and the mechanisms driving star formation quenching \citep[e.g.,][]{aung_entrainment_2024,park_rapid_2023,leung_mosdef_2019,li_mass-metallicity_2023}. This redshift range also encompasses the peak of AGN activity \citep[e.g.,][]{hopkins_cosmological_2008,davies_jwst_2024,park_widespread_2024}, making it a key window to study the interplay between black hole accretion and galaxy growth.

While rest-frame ultraviolet (UV) observations have provided critical insights into massive stars, dust attenuation, and the ionizing photon budget in high-redshift galaxies \citep[e.g.,][]{shivaei_mosdef_2018}, rest-frame optical emission lines remain the most direct tracers of many fundamental galaxy properties. Diagnostics based on [O~{\sc ii}], H$\beta$, [O~{\sc iii}], H$\alpha$, and [N~{\sc ii}] are widely used to determine star formation rates, gas-phase metallicities, excitation conditions, and AGN contribution \citep[e.g.,][also see e.g., \citealp{trouille_optx_2011,zeimann_hubble_2015,backhaus_clear_2022,backhaus_clear_2023,cleri_using_2023,trump_physical_2023,feuillet_classifying_2024,brinchmann_high-z_2023}]{kewley_understanding_2019, sanders_mosdef_2021, strom_nebular_2017, shapley_mosdef_2015}. However, these lines shift into the near-infrared (NIR) at $z\sim1$-3, where ground-based observations are challenged by atmospheric emission lines and strong telluric absorption and thermal emission that limit sensitivity, an issue particularly for obscured or faint sources.

The launch of \emph{JWST} has dramatically enhanced our ability to study galaxies during cosmic noon \citep[e.g.,][]{shapley_aurora_2025}. With its wide wavelength coverage and superb sensitivity, JWST's Near Infrared Spectrograph (NIRSpec) enables medium- to high-resolution spectroscopy across $0.6$-$5.3\,\mu$m, making it possible to capture the full suite of rest-optical lines for galaxies over $0.5\lesssim z \lesssim7$ \citep{jakobsen_near-infrared_2022}. Several large-scale NIRSpec programs have begun exploiting this capability. The JADES survey \citep{eisenstein_jades_2023, tacchella_jades_2023, rieke_marcia_data_2024} includes deep Micro-Shutter Array \citep[MSA,][]{ferruit_near-infrared_2022} spectroscopy in the GOODS fields, primarily targeting faint galaxies at $z>4$ \citep[][]{bunker_jades_2024,deugenio_jades_2025,curtis-lake_jades_2025,scholtz_jades_2025}. 
\added{RUBIES \citep{de_graaff_rubies_2025} targets rare, bright red sources and a reference galaxy sample over $1<z<10$ with PRISM and G395M.}
CEERS \citep{finkelstein_ceers_2023} and UNCOVER \citep{bezanson_jwst_2024} have extended the redshift frontier using prism-mode spectroscopy and lensing magnification. Meanwhile, GLASS \citep{treu_glass-jwst_2022} and CAPERS \citep{kokorev_capers_2025} have also emphasized galaxy properties during the epoch of reionization. Despite their breadth, most of these efforts prioritize either very high redshifts or low-resolution modes, leaving a gap in systematic, medium-resolution rest-optical spectroscopy for galaxies at cosmic noon \citep[see also e.g.,][]{maseda_nirspec_2024,slob_jwst-suspense_2024}.

In contrast, the Systematic Mid-Infrared Instrument Legacy Extragalactic Survey (SMILES) \citep{alberts_smiles_2024, rieke_smiles_2024} is uniquely positioned to study galaxy evolution during cosmic noon. SMILES combines eight-band MIRI imaging with medium-resolution NIRSpec spectroscopy over $\sim34$ arcmin$^2$ in the GOODS-S/HUDF region. The MIRI dataset offers continuous photometric coverage from 5.6 to 25.5 $\mu$m, enabling robust identification of dusty star-forming galaxies, obscured AGN, and mid-IR bright systems \citep[e.g.,][]{lyu_active_2024, shivaei_new_2024,williams_galaxies_2024,perez-gonzalez_what_2024}. Importantly, the SMILES field overlaps with some of the deepest available multi-wavelength ancillary data, including Chandra, HST, NIRCam, Spitzer, ALMA, and JVLA observations.

The SMILES NIRSpec component builds on this foundation by targeting $\sim$168 (166 were observed) galaxies spanning $0\lesssim z  \lesssim 7.5$, with a strong emphasis on galaxies at $z\sim1$-3 to better probe galaxy evolution during cosmic noon. The observations utilize the G140M/F100LP and G235M/F170LP gratings, providing continuous spectral coverage from $0.97$ to $3.07\,\mu$m at medium resolution ($R\sim1000$). This setup allows us to simultaneously detect key rest-frame optical emission lines such as [O~{\sc ii}], H$\beta$, [O~{\sc iii}], H$\alpha$, and [N~{\sc ii}] from z $\sim$ 1.6 to nearly 4, which are essential for measuring star formation rates, gas excitation, and \added{gas-phase} metallicity.

This paper presents the second SMILES data release, focused on the NIRSpec MSA component. We describe the sample selection, observing strategy, and data reduction process, and provide key data products including reduced 1D and 2D spectra, redshift measurements, and emission-line catalogs. The remainder of the paper is structured as follows: Section~\ref{sec:obs} details the target selection and observing setup. Section~\ref{sec:reduction} describes the data calibration and reduction pipeline. Section~\ref{sec:products} outlines the final data products, including spectral measurements and SED fits. Section~\ref{sec:science} demonstrates scientific applications of the dataset, including studies of AGN, outflows, and ionizing properties. Section~\ref{sec:summary} concludes with a summary and outlook for future work.

\section{Target Selection and Observation} \label{sec:obs}

\begin{figure*}[!ht]
    \centering
    \includegraphics[width=0.7\textwidth]{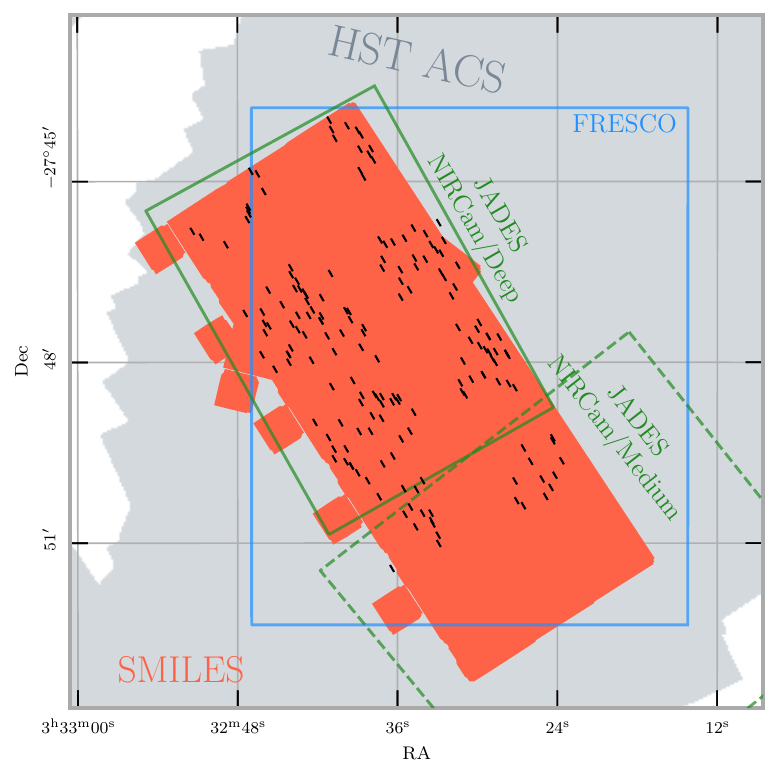}
    \caption{Footprint of the SMILES survey over the GOODS-S field, overlaid with key observational programs. The SMILES coverage is shown in orange, while the footprints of FRESCO (blue) and JADES (deep and medium tiers in  green) are also indicated. The black markers represent the locations of MSA slitlets that yield spectra from all three masks (size not to scale). The background shading provides a reference to the larger HST ACS footprint. This region is also covered by \textit{Chandra} X-ray observations, as well as JVLA radio observations at 3 GHz and 6 GHz, making it a prime multi-wavelength field for deep extragalactic studies. }
    \label{fig:footprint}
\end{figure*}

\begin{figure*}[!ht]
    \centering
    \includegraphics[width=0.9\linewidth]{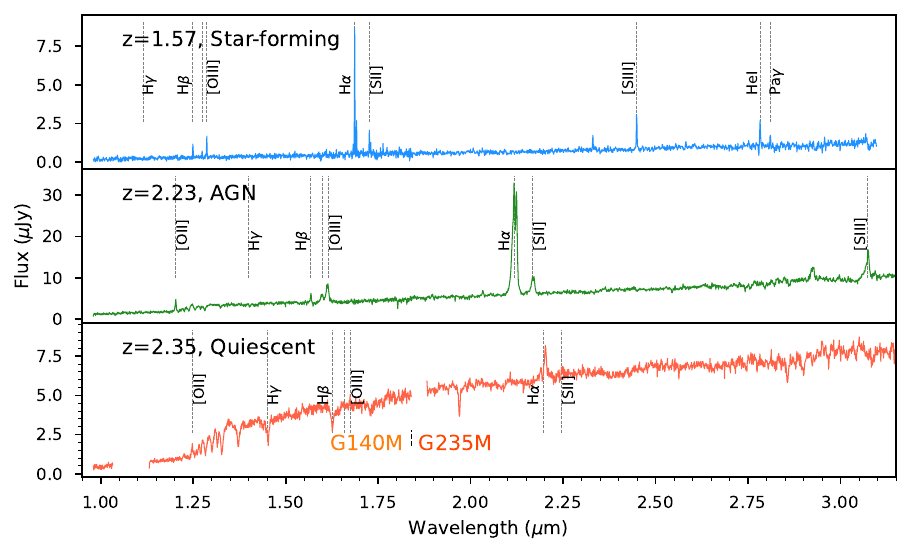}
    \caption{Example SMILES spectra from the G140M and G235M gratings. From top to bottom: a star-forming galaxy at $z = 1.57$ showing strong Balmer and forbidden emission lines; an AGN at $z = 2.23$ with prominent broad H$\alpha$ emission; and a quiescent galaxy at $z = 2.35$ dominated by  absorption features. Key emission/absorption lines are labeled. The small vertical dashed line near 1.8~$\mu$m marks the approximate boundary between the G140M and G235M grating coverage.}
    \label{fig:spec}
\end{figure*}

\subsection{Input Catalog}

The SMILES NIRSpec target selection was developed to build upon the extensive mid-infrared imaging provided by SMILES MIRI data, with a focus on galaxies during cosmic noon and a range of other key galaxy populations, 
\added{rather than following a single homogeneous analytic selection function}. 
Our selection strategy is designed to maximize the scientific value of the MSA observations by targeting galaxies with known dust, gas, and star formation properties. These include star-forming galaxies with varying PAH content, AGN candidates, and sources with multi-phase gas tracers from MIRI, ALMA, and NIRCam/grism surveys, 
\added{which we group into a set of science-motivated parent subsamples that are described in the following subsections}. 
The spectroscopic follow-up allows us to access rest-frame optical diagnostics critical for constraining ionization, metallicity, and extinction. The full input catalog leverages the rich multi-wavelength coverage in GOODS-S/HUDF, including deep imaging and spectroscopy from HST, JWST/NIRCam and NIRISS, ALMA, and VLA. Table \ref{tab:input} provides an overview of the input catalog. 

\subsubsection{Star-Forming Galaxies at Cosmic Noon}

A major science driver of SMILES is to understand the physical conditions in star-forming galaxies at $z\sim1$-2, the epoch when both star formation and black hole growth peaked. The availability of deep MIRI imaging in this redshift range enables detection of PAH emission, dust continuum, and warm dust SEDs. A key goal is to investigate how the PAH mass fraction relates to gas-phase metallicity. Local studies have established a strong dependence of PAH emission on metallicity \citep[see][]{li_spitzers_2020}, and recent SMILES-based work has shown that this trend persists out to $z\sim2$ when using stellar mass as a proxy \citep{shivaei_new_2024}. To directly calibrate this relation, we selected galaxies with a range of stellar masses for NIRSpec follow-up, enabling the use of strong-line metallicity diagnostics such as [O~{\sc iii}]/H$\beta$ and [N~{\sc ii}]/H$\alpha$ \citep[e.g.,][]{sanders_mosdef_2021}. 
\added{Candidates in this category are drawn from the SMILES initial data release catalog \citep{alberts_smiles_2024} within the MSA footprint, requiring robust MIRI detections and using JADES photometric redshifts together with {\tt Prospector}-based stellar masses and star formation rates \citep[][see \citealp{alberts_smiles_2024} for the detailed setup]{johnson_stellar_2021}; we do not impose any additional color cuts beyond these catalog selections, so that a broad range of star-forming galaxies is included.}

In addition, we targeted sources where Paschen lines can be observed to constrain dust attenuation in a complementary fashion. The SMILES field overlaps with the FRESCO NIRCam grism survey \citep{oesch_jwst_2023}, which captures Pa$\alpha$ and Pa$\beta$ in the NIRCam/F444W grism. When combined with NIRSpec spectroscopy covering Balmer lines, these targets allow us to test nebular attenuation curves by comparing Paschen and Balmer recombination line ratios and their relation to MIRI-measured warm dust emission \citep[e.g.,][]{reddy_jwstaurora_2025}.

Finally, we include galaxies with deep ALMA coverage. The GOODS-S field contains legacy ALMA datasets such as ASPECS \citep{walter_alma_2016} and ASAGAO \citep{ueda_alma_2018}, along with other programs targeting submillimeter galaxies and line emitters \citep{boogaard_midis_2024, dunlop_deep_2017, adscheid_a3cosmos_2024}. Many of these galaxies are detected across multiple MIRI bands and have Paschen line coverage from FRESCO. The combination of MIRI, ALMA, and NIRSpec allows us to jointly study PAH strength, extinction, metallicity, and cold gas content. This multi-phase perspective offers new insights into the interplay of dust, gas, and metals in galaxies across the main sequence at cosmic noon \citep{popesso_main_2023}.

\subsubsection{Obscured AGN across Cosmic Time}

The SMILES dataset is also designed to search and characterize the obscured AGN population and improve the AGN census combined with other multi-wavelength datasets in the central region of GOODS-S field (e.g, \citealp{sun_extreme_2025, rieke_confirming_2025}). Based on comprehensive SED analysis of HST, JWST/NIRCam and JWST/MIRI photometry, 
\citet{lyu_active_2024} identified a substantial sample of AGN candidates in the MIRI bands, many of which show evidence of heavy dust obscuration. 
\added{Our obscured AGN parent sample is drawn directly from this catalog, and we adopt their AGN classifications without imposing any additional color cuts.}
NIRSpec spectroscopy allows us to probe the AGN evidence in the rest-frame optical wavelengths and characterize key physical properties. For AGN at $z\sim1$-3.7, we simultaneously cover H$\alpha$, H$\beta$, [O~{\sc iii}], and [N~{\sc ii}], enabling rest-frame optical classification via diagnostic diagrams such as the BPT \citep{baldwin_classification_1981}. For broad-line AGN, we can also derive virial black hole masses using the line widths of H$\alpha$, H$\beta$, and Paschen lines at lower redshift. Additionally, emission line profiles in these sources allow us to search for outflow and inflow signatures associated with AGN feedback. 

\subsubsection{Quiescent Galaxies and Other High-redshift Targets}

In addition to the primary science categories above, the SMILES NIRSpec masks include several other galaxy types to broaden the scientific return. We include massive quiescent galaxies at $z>1$, \added{drawn from the JADES quiescent catalog of} \citep{ji_jades_2024}, to investigate post-starburst and aging stellar populations. Candidate members of protoclusters and overdense structures at $z\sim2$ and 5 are also included, based on photometric redshifts and environmental selection \added{from} \citep[][P.~Rinaldi et al.~in prep.]{helton_jwst_2023}. We further target emission-line galaxies identified in the FRESCO grism data at $z>3$ (\citealt{hainline_cosmos_2024} and F.\ Sun et al.\ private communication), as well as radio sources from deep 6 GHz VLA imaging at $\sim0.45''$ resolution \citep{alberts_completing_2020}. These filler targets provide opportunities to explore a wide range of galaxy environments, gas excitation conditions, and evolutionary stages.

\subsection{Observation Setup}

The SMILES NIRSpec observations were carried out in multi-object spectroscopy (MOS) mode using NIRSpec/MSA. We designed three MSA masks with customized spectroscopic configurations to cover our science field in GOODS-S. For each mask, we observed with two medium-resolution grating/filter combinations: G140M/F100LP and G235M/F170LP. This setup provides continuous spectral coverage from $0.97$ to $3.17\,\mu$m at a resolution of $R \sim 1000$, sufficient to deblend closely spaced emission lines such as H$\alpha$ and [N~{\sc ii}].

The selected grating configuration allows access to key rest-frame optical emission lines across a broad redshift range: [O~{\sc ii}] $\lambda\lambda3726,3729$ for $1.6 < z < 7.5$, H$\beta$ and [O~{\sc iii}] $\lambda\lambda4959,5007$ for $1.1 < z < 5.3$, H$\alpha$ and [N~{\sc ii}] $\lambda6584$ for $0.5 < z < 3.7$.
We also cover Paschen-series hydrogen lines, including Pa$\alpha$ at $z < 0.7$ and Pa$\beta$ at $z < 1.45$, which provide independent constraints on nebular dust attenuation when combined with Balmer lines \citep{calzetti_dust_2000}.

All observations used a three-shutter slitlet configuration combined with a three-nod dither pattern to improve background subtraction and mitigate the impact of detector artifacts. For each grating, we employed an exposure setup consisting of 16 groups per integration and two integrations with the {\tt NRSIRS2} readout mode, leading to a total of 3.89 hours of science exposure time per mask.

The mask design was optimized using the {\tt eMPT} software \citep{bonaventura_near-infrared_2023}, which offers advanced planning capabilities beyond the standard MSA Planning Tool (MPT\footnote{\url{https://jwst-docs.stsci.edu/jwst-near-infrared-spectrograph/nirspec-apt-templates/nirspec-multi-object-spectroscopy-apt-template/nirspec-msa-planning-tool-mpt}}
) in the JWST Astronomer’s Proposal Tool (APT\footnote{\url{https://jwst-docs.stsci.edu/jwst-astronomer-s-proposal-tool-overview}}
). eMPT enables flexible prioritization across science categories while optimizing slit allocation through the use of an algorithm that maximizes the number of high-priority targets included in a single MSA observation. 
\added{In practice, we supplied eMPT with the parent catalogs described above, together with a discrete set of priority flags that encode the relative importance of each science category}. 
We assigned the highest priority to rare or high-value sources, including AGN candidates and star-forming galaxies with ancillary MIRI, ALMA, or FRESCO Paschen line data; remaining sources were assigned to lower priority levels (2 through 6) based on science relevance, number density, and spatial distribution.
\added{Whether a given candidate is ultimately allocated a slit is therefore determined by its assigned priority and its position relative to neighboring candidates in the MSA, not by any additional internal ranking within eMPT; for a fixed priority level, candidates in crowded regions have lower average targeting fractions than those in sparse regions}.

Through extensive testing of various priority weightings and slit allocation schemes, we converged on the three mask configurations that maximized the number of high-priority targets observed. Across the three masks, slits were allocated to a total of 168 unique sources (166 were observed). Among these, approximately 40\% fall into the star-forming galaxy category and 30\% are classified as AGN candidates, with the remainder comprising other science targets including high-redshift emission-line galaxies, quiescent galaxies, and radio sources. 
\added{Tables \ref{tab:input} and \ref{tab:output} summarize how many objects in each parent category received slits: for example, we placed slits on 70 of 464 cosmic noon star-forming candidates (about 15\%), 50 of 382 AGN candidates (about 13\%), 40 of 447 FRESCO overdensity galaxies (about 9\%), 28 of 209 radio or ALMA sources (about 13\%), 9 of 63 Paschen+MIRI systems (about 14\%), 13 of 49 quiescent galaxies (about 27\%), 4 of 24 high-redshift FRESCO emitters (about 17\%), and 2 of 22 other galaxies (about 9\%). Because many galaxies satisfy more than one criterion, the sum of the category counts exceeds the number of unique targets}.
Figure \ref{fig:footprint} shows our survey footprint along with the sky positions of micro-shutters. Example spectra from different types of galaxies in the SMILES sample are shown in Figure \ref{fig:spec}.

\begin{table*}[]
    \addtocounter{table}{-1}
    \centering
    \begin{tabular}{lccccccccr}
\toprule
label & cosmicnoon sf & agn & fresco overdensity & radio/alma & paschen miri & qgs & fresco highz & other & count \\
\hline
 &  &  & x &  &  &  &  &  & 418 \\
 & x &  &  &  &  &  &  &  & 322 \\
 &  & x &  &  &  &  &  &  & 304 \\
 &  &  &  & x &  &  &  &  & 145 \\
 & x &  &  &  & x &  &  &  & 43 \\
 & x &  &  & x &  &  &  &  & 36 \\
 & x & x &  &  &  &  &  &  & 35 \\
 &  &  &  &  &  & x &  &  & 24 \\
 &  & x &  & x &  &  &  &  & 20 \\
 &  &  & x &  &  &  & x &  & 20 \\
 &  &  &  &  &  &  &  & x & 18 \\
 & x &  &  &  &  & x &  &  & 18 \\
 &  & x &  &  & x &  &  &  & 9 \\
 &  & x & x &  &  &  &  &  & 6 \\
 &  &  &  &  &  &  & x &  & 4 \\
 &  &  &  &  & x &  &  &  & 4 \\
 &  &  & x &  &  &  &  & x & 3 \\
 & x &  &  & x &  & x &  &  & 3 \\
 & x & x &  &  & x &  &  &  & 3 \\
 &  & x &  &  &  & x &  &  & 2 \\
 &  & x &  &  &  &  &  & x & 1 \\
 &  &  &  & x & x &  &  &  & 1 \\
 & x &  &  &  & x & x &  &  & 1 \\
 &  &  &  & x &  & x &  &  & 1 \\
 & x &  &  & x & x &  &  &  & 1 \\
 & x & x &  & x &  &  &  &  & 1 \\
 & x & x &  & x & x &  &  &  & 1 \\
Total & 464 & 382 & 447 & 209 & 63 & 49 & 24 & 22 & 1444 \\
\hline
\end{tabular}
    \caption{Summary of unique keyword combinations present in the input sample. Each row represents a distinct combination of high-level categories (e.g., \texttt{agn}, \texttt{cosmicnoon sf}, \texttt{radio/alma}) identified in individual sources. The total row shows the number of sources associated with each keyword across all combinations.}
    \label{tab:input}
\end{table*}

\begin{table*}[]
    \centering
    \begin{tabular}{lccccccccr}
\toprule
label & cosmicnoon sf & agn & fresco overdensity & radio/alma & paschen/miri & qgs & fresco highz & other & count \\
\hline
 & x &  &  &  &  &  &  &  & 38 \\
 &  &  & x &  &  &  &  &  & 36 \\
 &  & x &  &  &  &  &  &  & 29 \\
 & x & x &  &  &  &  &  &  & 12 \\
 & x &  &  & x &  &  &  &  & 9 \\
 &  &  &  & x &  &  &  &  & 8 \\
 &  &  &  &  &  & x &  &  & 6 \\
 &  & x &  & x &  &  &  &  & 6 \\
 & x &  &  &  & x &  &  &  & 5 \\
 &  &  & x &  &  &  & x &  & 4 \\
 & x &  &  & x &  & x &  &  & 3 \\
 &  &  &  &  &  &  &  & x & 2 \\
 &  &  &  &  & x &  &  &  & 2 \\
 & x &  &  &  &  & x &  &  & 2 \\
 &  & x &  &  & x &  &  &  & 1 \\
 &  & x &  &  &  & x &  &  & 1 \\
 &  &  &  & x &  & x &  &  & 1 \\
 & x & x &  & x & x &  &  &  & 1 \\
Total & 70 & 50 & 40 & 28 & 9 & 13 & 4 & 2 & 166 \\
\hline
\end{tabular}
    \caption{Summary of unique keyword combinations present in the observed sample. }
    \label{tab:output}
\end{table*}

\section{Data Reduction} \label{sec:reduction}

The SMILES NIRSpec data are processed using the {\tt JWST Calibration Pipeline} \citep{bushouse_jwst_2022}, version 1.14.0, with the Calibration Reference Data System (CRDS) mapping jwst\_1236.pmap\footnote{\url{https://jwst-crds.stsci.edu/}}. The reduction follows a three-stage approach: detector-level corrections, spectroscopic calibration and background subtraction, and spectral extraction and combination. Each stage incorporates custom processing steps to optimize data quality and mitigate systematic effects.

\subsection{Stage 1: Detector-Level Processing}

\begin{figure*}[!ht]
    \centering
    \includegraphics[width=0.8\linewidth]{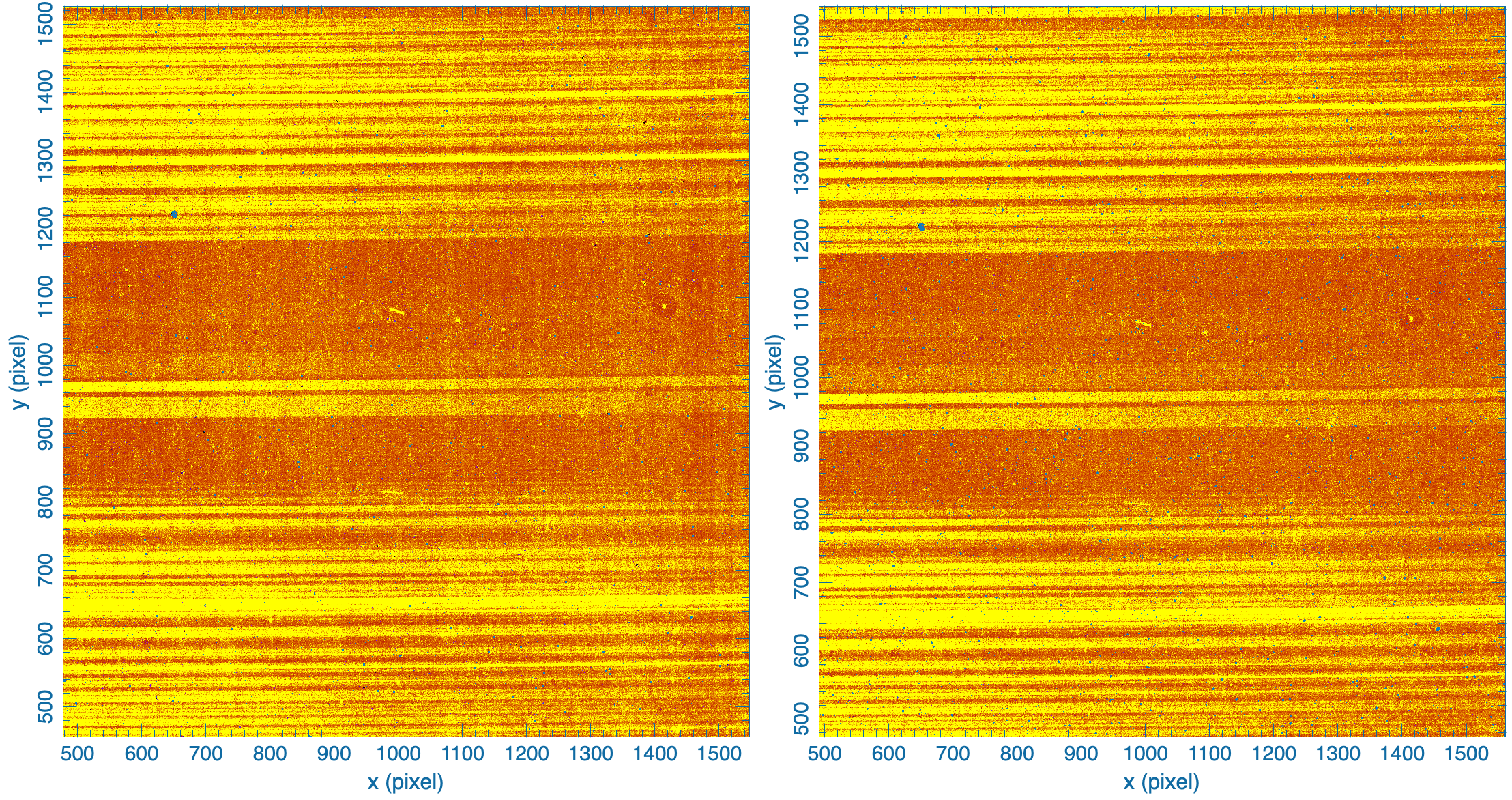}
    \caption{Correction of 1/$f$ noise in NIRSpec 2D spectra. \textbf{Left:} Original uncorrected image showing prominent low-frequency vertical banding due to correlated readout noise. \textbf{Right:} Same exposure after applying the 1/$f$ correction using a Chebyshev polynomial fit along detector columns. The correction significantly reduces the striping pattern and improves the background uniformity across the detector.}
    \label{fig:1f}
\end{figure*}

The raw uncalibrated exposures undergo initial processing to remove instrumental artifacts. Reference pixel correction is applied to account for detector drifts, followed by superbias subtraction and dark current removal. Cosmic ray events are flagged and rejected using the {\tt jump} detection algorithm, with a rejection threshold of 3$\sigma$. To address bright transient events (``snowballs'') that affect the NIRSpec detectors, we expand the jump flagging to neighboring pixels while setting a minimum detection area of 10 pixels to reduce false detections. The post-jump flagging threshold is set to 1000 DN, with a temporal cutoff of 10 seconds to avoid erroneously flagging extended sources.

To further refine the data, a custom bad pixel rejection method is applied after Stage 1. Hot and bad pixels are identified following an approach adopted from MIRI bad pixel flagging, which evaluates spatial variations across the detector \footnote{\url{https://github.com/STScI-MIRI/MRS-ExampleNB/blob/main/D2P_Notebooks/MRS_Flag_Badpix.ipynb}}. The method calculates the first-order difference along the detector rows and columns, applying a median filter with a kernel size of $5 \times 5$ pixels to create a normalized difference image. A percentile-based threshold is then used to flag outliers. Pixels exceeding the 99.95th percentile of the normalized difference in the central region (excluding detector edges) are flagged, with an additional mask expansion applied to include adjacent pixels in the four cardinal directions. To ensure robustness, a secondary thresholding step at the 99.7th percentile is performed to capture more subtle defects. The final bad pixel mask is incorporated into subsequent processing steps to prevent contamination in extracted spectra.

An additional correction is applied to mitigate $1/f$ noise arising from the readout electronics. This manifests as low-frequency striping along detector columns and rows. We use a Chebyshev polynomial fit to model and subtract the $1/f$ background, as implemented in the {\tt exposure\_oneoverf\_correction} function of {\tt grizli} \citep{brammer_grizli_2019}. The correction is performed along detector columns (axis = 0), with a polynomial degree scaling factor of 256 pixels per degree to effectively capture the low-frequency variations without overfitting. This approach ensures that the correction removes systematic trends while preserving real astrophysical signal, significantly improving the uniformity of the spectral background and reducing correlated noise that would otherwise degrade the extracted spectra. Figure \ref{fig:1f} shows an example of our correction.

\subsection{Stage 2: Spectroscopic Calibration and Background Subtraction}

After correcting detector-level artifacts, the data undergo spectroscopic calibration. Wavelength solutions are derived using internal lamp reference spectra, enabling accurate mapping of detector pixels to physical wavelengths. Flat-field corrections are applied using calibration files from the CRDS system to account for pixel-to-pixel sensitivity variations and wavelength-dependent throughput. Flux calibration is then performed by scaling the extracted spectra based on JWST standard star observations. To verify the wavelength calibration, we compared the redshifts of SMILES sources to those of overlapping FRESCO detections \citep{meyer_jwst_2024}, including 10 H$\alpha$ emitters and 2 [O~{\sc iii}] emitters at $z > 5$. We find a mean velocity offset of $-103$~km~s$^{-1}$ between the two samples, which is within the expected uncertainty given the NIRSpec medium-resolution mode ($R\sim1000$, corresponding to $\Delta v \sim 300$~km~s$^{-1}$) and a similar error in the wavelength calibration of the NIRCam grism \footnote{\url{https://jwst-docs.stsci.edu/jwst-calibration-status}}\citep{sun_first_2023}. We  do not attempt to correct the wavelength solution from the uneven source light distribution within the shutter.

\begin{figure*}[!ht]
    \centering
    \includegraphics[width=0.9\linewidth]{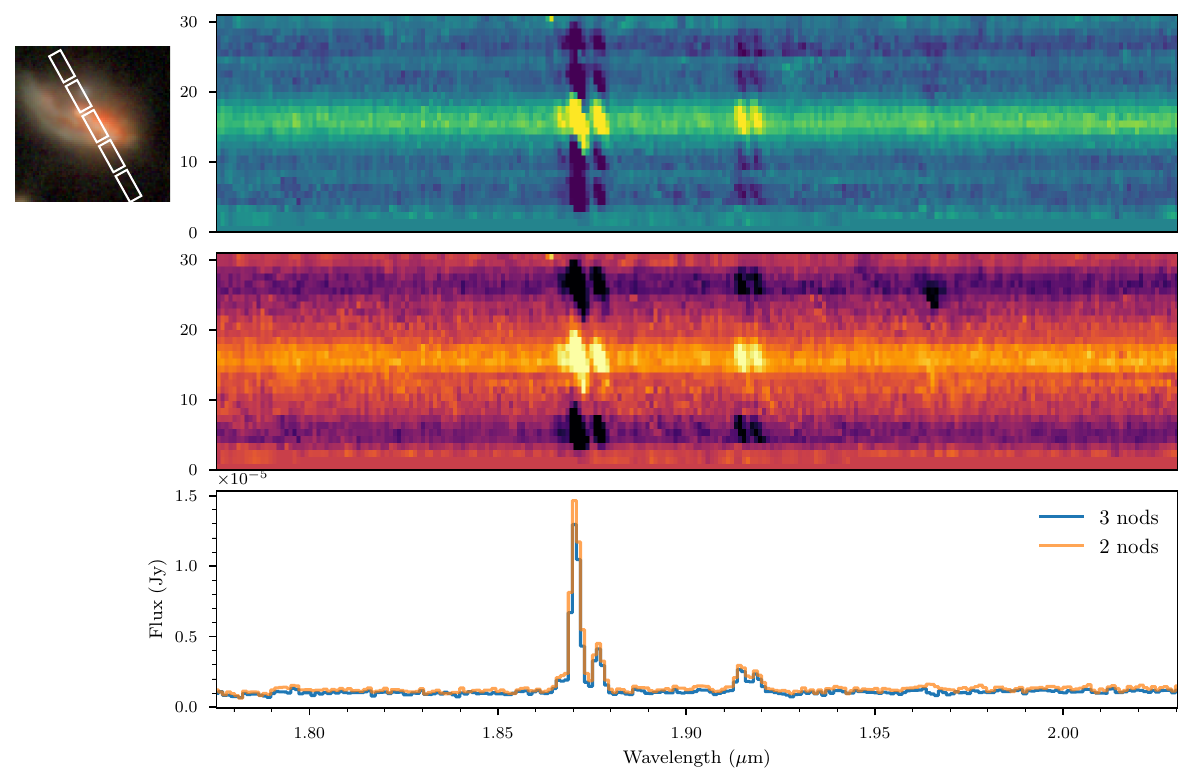}
\caption{Comparison between standard 3-nod (top) and modified 2-nod (middle) extractions for a spatially extended source in the SMILES dataset (ID: 199580, $z=1.8501$). The top panel includes background subtraction using all three nod positions, which can partially suppress extended emission due to self-subtraction. In contrast, the middle panel excludes the central nod during background subtraction and stacking, preserving more of the extended flux. The 2D and 1D spectra (bottom) show a stronger H$\alpha$ peak under the 2-nod scheme. The NIRCam RGB image (F200W/F150W/F115W), with the slit position overplotted, is shown at upper left.}
    \label{fig:2nod}
\end{figure*}

Accurate local sky background subtraction is essential for optimizing the detection of faint emission lines. For compact targets, we employ a three-shutter nodding pattern, in which the background at each nod position is estimated from the other two exposures where the source is absent. This approach effectively removes residual background fluctuations and scattered light. However, for spatially extended sources, this method can result in self-subtraction of diffuse emission. In these cases confirmed by visual inspection
\added{of both the NIRCam imaging and the NIRSpec 2D spectra}
, we instead adopt a two-shutter nodding scheme, excluding the central nod from both background subtraction and co-addition. Using only the $+1$ and $-1$ nod positions allows us to preserve extended line emission. This modified approach has also been successfully used in other NIRSpec studies, such as JADES \citep{deugenio_jades_2025}. Figure~\ref{fig:2nod} illustrates the difference between 2-nod and 3-nod reductions for an extended galaxy in the SMILES dataset.

\subsection{Stage 3: Spectral Extraction and Combination}

Finally, the exposures are aligned and resampled onto a common grid and we use a standard 6-pixel extraction window to extract 1d spectra using {\tt calwebb\_spec3}. One caveat is that, multiple works have noted that the pipeline may underestimate the error on flux compared to the pixel-to-pixel variation \citep[e.g.,][]{brinchmann_high-z_2023,trump_physical_2023,maseda_jwstnirspec_2023}, and there should be a correction factor of $1.7\sim1.9$. However, here we just keep the original error array from the pipeline and users can re-scale the error array as needed.

\begin{figure*}[!ht]
    \centering
    \gridline{\fig{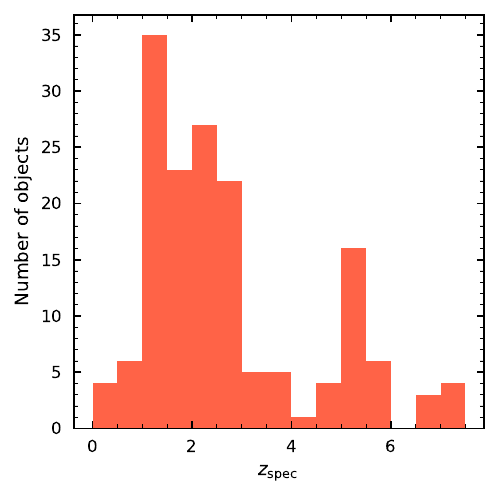}{0.4\textwidth}{(a)}
          \fig{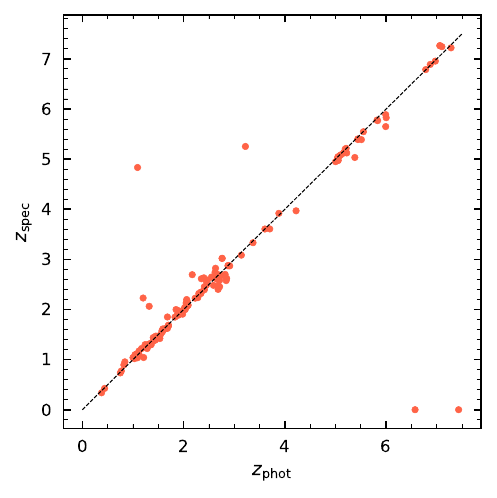}{0.4\textwidth}{(b)}}
    \caption{Summary of redshift measurements in the SMILES NIRSpec sample. \textbf{(a)} Histogram of confirmed spectroscopic redshifts ($z_{\rm spec}$), showing broad coverage from $z < 1$ to $z > 6$, with a peak near $z\sim2$. \textbf{(b)} Comparison between spectroscopic redshifts and photometric redshifts ($z_{\rm phot}$) derived from multi-band imaging. The dashed line shows the one-to-one relation. Most sources lie close to the line, demonstrating good agreement, with a small number of outliers primarily at high redshift.}
    \label{fig:specz}
\end{figure*}

\begingroup
\setlength{\tabcolsep}{11pt}
\begin{deluxetable}{ccccccccc}
\tablenum{3}
\tabletypesize{\footnotesize}
\tablewidth{0pt}
\tablecaption{SMILES Sources with MSA spectroscopy\label{tab:final_spectra}}
\tablehead{
\colhead{NIRCam ID} & \colhead{MIRI ID} & \colhead{RA (deg)} & \colhead{DEC (deg)} & \colhead{$z_{\mathrm{photo}}$} & \colhead{$z_{\mathrm{spec}}$} & \colhead{Q} & \colhead{AGN?} & \colhead{Extended?}}
\decimalcolnumbers
\startdata
114544 &  & 53.183350 & -27.796024 & 5.00 & 4.9485 & A &  &  \\
124087 & 1184 & 53.178243 & -27.783152 & 1.09 & 1.1204 & A &  & Y \\
124908 & 1205 & 53.148982 & -27.781990 & 1.88 & 1.9068 & A &  & Y \\
139052 & 1450 & 53.211246 & -27.765383 & 1.70 & 1.6862 & A &  &  \\
154775 & 3010 & 53.169570 & -27.738061 & 7.10 & 7.2432 & C & Y &  \\
155734 &  & 53.165762 & -27.734578 & 5.43 &  & D &  &  \\
201793 & 947 & 53.188277 & -27.801935 & 4.03 &  & D &  &  \\
202889 & 1001 & 53.192359 & -27.797915 & 1.17 & 1.2208 & A &  &  \\
203038 &  & 53.184045 & -27.797831 & 7.06 & 7.2627 & C &  &  \\
204330 &  & 53.179001 & -27.792428 & 5.21 &  & D &  &  \\
204467 & 1084 & 53.191341 & -27.791815 & 2.07 & 2.1642 & A &  & Y \\
204621 & 1095 & 53.181166 & -27.790993 & 2.62 & 2.7404 & C &  &  \\
204949 & 1108 & 53.190219 & -27.789925 & 2.59 & 2.4808 & A &  & Y \\
205123 & 1118 & 53.173891 & -27.788570 & 1.70 & 1.6736 & A &  &  \\
205144 &  & 53.183009 & -27.789457 & 6.97 & 6.9547 & D &  &  \\
205194 & 1113 & 53.191562 & -27.789249 & 2.69 & 2.5892 & A &  &  \\
205818 & 1146 & 53.197600 & -27.786474 & 1.12 & 1.0984 & B &  & Y \\
206144 & 1152 & 53.192244 & -27.786084 & 2.71 & 2.4541 & A &  & Y \\
206183 & 1155 & 53.176590 & -27.785521 & 1.34 & 1.3171 & A &  & Y \\
206924 & 1177 & 53.186048 & -27.784080 & 2.41 & 2.3941 & A &  &  \\
\enddata
\tablecomments{Columns: (1) JADES NIRCam identifier; (2) SMILES MIRI identifier; (3) Right Ascension in degrees (J2000); (4) Declination in degrees (J2000); (5) Photometric redshift; (6) Spectroscopic redshift; (7) \added{Spectroscopic redshift quality flag (A, B, C, or D; see text for definitions);} (8) Flag indicating AGN classification based on \citet{rieke_confirming_2025}; (9) Flag indicating whether the source is extended and therefore extracted with two nodding positions. The complete table is available in FITS format on MAST.}
\end{deluxetable}
\endgroup

\section{Data Products}\label{sec:products}

\subsection{Reduced Spectra}

The final data products include both 2D and 1D spectra for all observed galaxies. For each source, we also include metadata indicating the slit position angle (PA, in degrees) and the source location within the shutter, expressed as fractional coordinates along the x and y axes (source\_xpos and source\_ypos), where a value of 0.5 corresponds to the shutter center. All 1D extractions are performed using either a three-nod or two-nod background subtraction scheme, depending on the source morphology.

To account for the loss of extended flux outside the micro-shutter aperture, we provide an optional set of empirically slit-loss-corrected spectra. While the default pipeline applies corrections assuming point-like sources, this approximation breaks down for galaxies with significant spatial extent. We address this by computing wavelength-dependent corrections using photometry from multiple NIRCam filters. For most targets, we adopt F115W, F150W, and F210M to minimize contamination and avoid detector gap regions. For sources requiring special care, e.g., those near detector edges or with irregular morphologies, we include additional filters such as F182M, F200W, and F277W. The correction factor is computed as the ratio of the flux within the MSA shutter to the Kron-convolved total flux in each band, and a second-order polynomial is fit to model the wavelength dependence.

Each slit-loss correction is visually validated to ensure consistency and to avoid biases from photometric contamination or aperture mis-centering. The resulting correction curve is then applied to the extracted 1D spectra. To verify the accuracy of this correction, we compare synthetic photometry generated from the corrected spectra with broadband photometric measurements. This procedure yields flux corrections with a typical uncertainty of $\pm 0.2$ dex and a median offset of zero, confirming that the method effectively recovers the total galaxy light. The corrected spectra are suitable for line luminosity measurements; however, for line ratio analyses (e.g., BPT diagram) we recommend using the uncorrected spectra to avoid introducing wavelength-dependent systematics from the correction process.

\subsection{Redshift Catalog}

The spectroscopic redshifts ($z_{\rm spec}$) for the SMILES NIRSpec sample are determined primarily from emission lines for star-forming galaxies\added{, using automated fits with GELATO}. \added{These fits typically include rest-frame optical and near-infrared features such as [O~{\sc ii}] $\lambda\lambda3726,3729$, H$\beta$, [O~{\sc iii}] $\lambda\lambda4959,5007$, H$\alpha$, [N~{\sc ii}] $\lambda6584$, [S~{\sc ii}] $\lambda\lambda6717,6731$, [S~{\sc iii}] $\lambda\lambda9069,9531$, He~{\sc i}, and Pa$\gamma$. Each redshift is then visually inspected by at least two team members to confirm the automated solution and, where necessary, fine-tune the line centroids and adopted $z_{\rm spec}$.}

For quiescent galaxies, where emission lines are weak or absent, redshifts are derived from stellar absorption features, including the Ca~{\sc ii} H \& K doublet, Balmer absorption lines, and Mg~{\sc i} $\lambda5175$\added{, with the primary $z_{\rm spec}$ measurement coming from {\tt pPXF} fits to the full absorption-line spectrum}. \added{The same visual inspection procedure is applied to verify the {\tt pPXF} solution and apply small manual adjustments when warranted.} When both emission and absorption features are available, we cross-validate the redshift estimates to check for agreement.

To assess redshift accuracy, in addition to comparing our redshifts with FRESCO (see Section \ref{sec:obs}), 12 targets fall in the NIRISS NGDEEP footprint in HUDF and the independent crosscheck reveals 0 outliers (P.~Rinaldi et al.~in prep). 
\added{The typical offset between our spectroscopic redshifts and the slitless values is $\langle \Delta z / (1+z) \rangle \approx -0.0002$ with a standard deviation of $\sigma[\Delta z / (1+z)] \approx 0.0002$.}
We also compare our spectroscopic redshifts with photometric redshifts ($z_{\rm phot}$) obtained from multi-band imaging using the EAZY code \citep{brammer_eazy_2008} presented in the JADES catalog \citep{eisenstein_jades_2023}. As shown in Figure~\ref{fig:specz}, there is strong overall agreement between $z_{\rm spec}$ and $z_{\rm phot}$, with the majority of sources lying close to the one-to-one line. The redshift distribution of the final sample is also displayed, highlighting broad redshift coverage with a concentration near $z \sim 2$. Table \ref{tab:final_spectra} presents our final catalog of spectra with redshifts listed. \added{We assign a redshift quality flag {\tt Q} with four values: A, B, C, and D. A indicates secure redshifts based on multiple high S/N features, B indicates moderate confidence (typically a single strong or broad line with supporting evidence), C indicates low-confidence redshifts, and D indicates that no reliable spectroscopic redshift can be determined.} \added{For reference, the GELATO and {\tt pPXF} redshifts are listed in Tables \ref{tab:gelato_example} and \ref{tab:ppxf_example}, and are also included in the released FITS tables.} 

\subsection{Emission Line Catalog}

\begin{figure*}[!ht]
    \centering
    \includegraphics[width=1\linewidth]{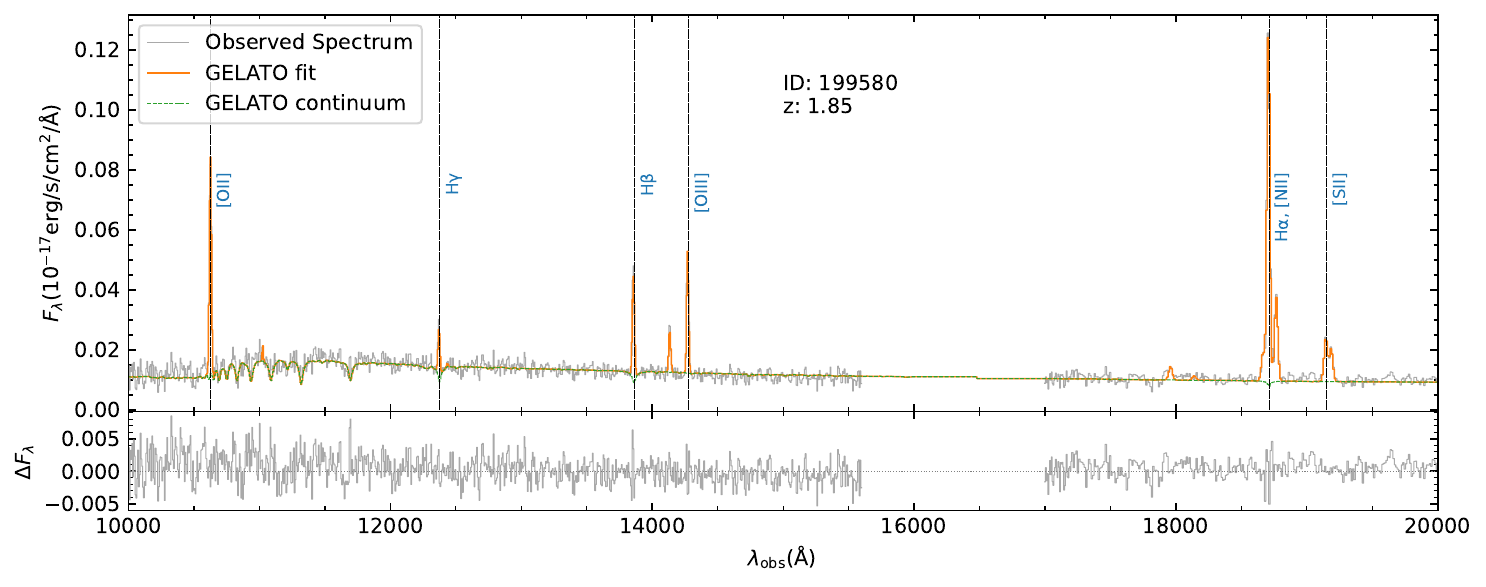}
    \includegraphics[width=1\linewidth]{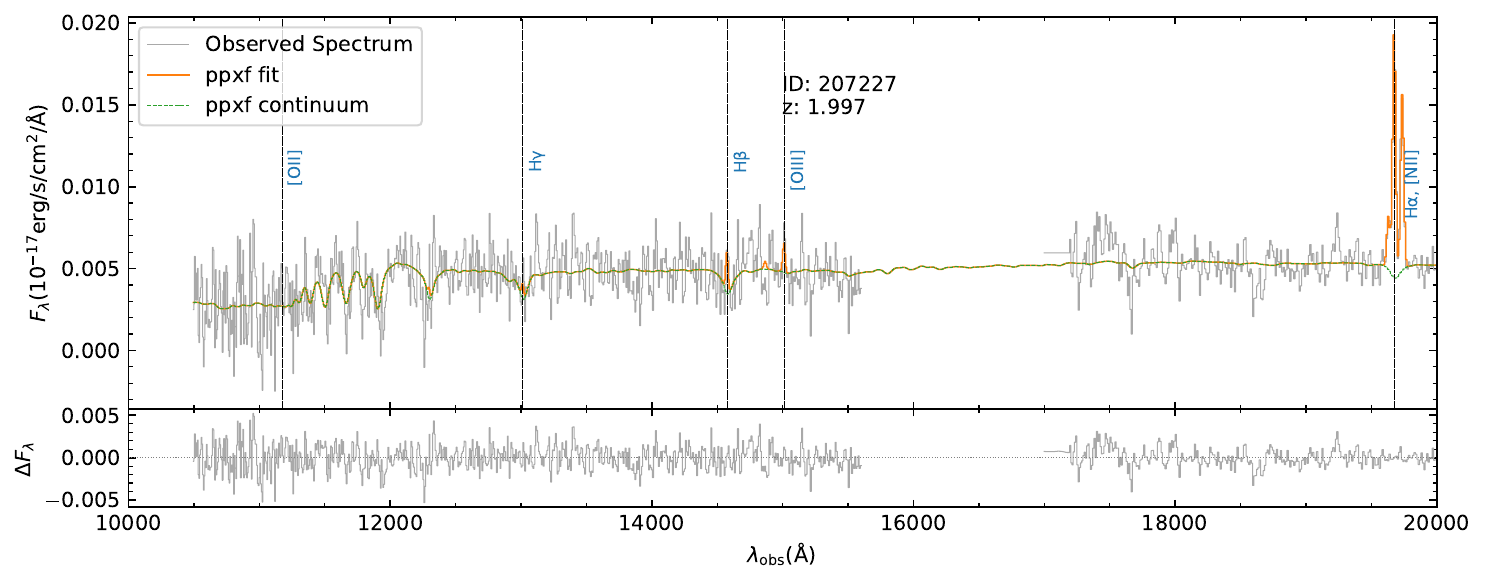}
    \caption{\added{Examples of our spectral fitting with GELATO (top) and {\tt pPXF} (bottom). In each case the observed spectrum is shown in gray, the continuum model in green (dotted), and the total best fit model in orange (solid). The lower panels show the residuals after subtracting the best fit model, and vertical dashed lines mark the  positions of key rest-optical emission lines.}}
    \label{fig:fit}
\end{figure*}

The emission-line properties of the SMILES NIRSpec sample are mainly measured using the GELATO software\footnote{\url{https://github.com/TheSkyentist/GELATO}} \citep{hviding_theskyentistgelato_2022}, which provides a robust framework for modeling both the stellar continuum and emission lines. GELATO employs a combination of spectral population synthesis and Gaussian emission-line fitting, allowing for a self-consistent decomposition of galaxy spectra.
Briefly, GELATO uses the Extended MILES (E-MILES) stellar population synthesis models\footnote{\url{http://research.iac.es/proyecto/miles/}}, assuming a Chabrier initial mass function (IMF) \citep{chabrier_galactic_2003}. The templates span a wide range of metallicities and ages, ensuring an accurate reconstruction of the underlying stellar continuum. The emission lines are fitted as Gaussians, with free parameters for line flux, velocity dispersion, and redshift. The fitting process includes a continuum-subtraction step to isolate emission lines from the stellar background, improving the accuracy of line flux measurements.

For sources with no significant emission lines, such as QGs, GELATO does not perform well, as the continuum dominates the spectrum. Instead, we use the pPXF method \citep{cappellari_ppxf_2012,cappellari_full_2023}, which is optimized for stellar continuum modeling. The stellar templates for pPXF are built using the Flexible Stellar Population Synthesis (FSPS) model \citep{conroy_propagation_2009, conroy_propagation_2010}, with the MILES spectral library \citep{sanchez-blazquez_medium-resolution_2006, falcon-barroso_updated_2011}, the MIST isochrones \citep{choi_mesa_2016}, and a Salpeter IMF \citep{salpeter_luminosity_1955}. Even though the Salpeter IMF differs from the Kroupa IMF commonly used for SED fitting, pPXF results are largely insensitive to the low-mass IMF slope \citep{cappellari_full_2023}.

The FSPS-based templates cover a wide range of stellar ages (1 Myr to 15.85 Gyr) and metallicities ([Z/H]$=-1.75$ to $0.25$) in logarithmic steps of 0.1 dex for age and 0.25 dex for metallicity. To ensure physical consistency, we limit the template age to be no greater than the Universe’s age at the redshift of each galaxy. During pPXF fitting, we mask prominent nebular emission lines, including Balmer series lines (H$\beta$, H$\alpha$), forbidden lines such as [O~{\sc iii}] $\lambda\lambda4959,5007$, [O~{\sc i}] $\lambda6300$, [N~{\sc ii}] $\lambda\lambda6548,6584$, and [S~{\sc ii}] $\lambda\lambda6717,6731$. After subtracting the best-fit stellar continuum, we perform an independent Gaussian fitting of these emission lines to obtain accurate flux and kinematic measurements.

To ensure robust measurements, all emission-line fits are visually inspected, and fluxes are cross-checked against previous spectroscopic datasets where available. We note that these fits are performed on spectra without slit-loss correction to avoid any additional bias. The final emission-line catalog provides line fluxes, equivalent widths, line redshift, etc., facilitating studies of ionized gas conditions, star formation rates, and AGN activity across cosmic time. Table \ref{tab:gelato_example} and Table \ref{tab:ppxf_example} provide examples of our GELATO and pPXF fit results, respectively, \added{and Figure \ref{fig:fit} exemplifies the corresponding spectral fits.}

\begingroup
\setlength{\tabcolsep}{12pt}
\begin{deluxetable*}{lrrrrrrr}
\tablenum{4}
\tabletypesize{\footnotesize}
\tablewidth{0pt}
\tablecaption{GELATO Emission-Line Fit Results\label{tab:gelato_example}}
\tablehead{
\colhead{NIRCam ID} & 
\colhead{$z_{\rm SSP}$} & 
\colhead{H$\alpha$ Flux} & 
\colhead{Flux Err} & 
\colhead{REW} & 
\colhead{REW Err} & 
\colhead{Line Redshift} & 
\colhead{Redshift Err}
}
\decimalcolnumbers
\startdata
207982 & 1.9983 & 0.556 & 0.015 & 112.88 & 3.08 & 1.9977 & \nodata \\
204621 & 2.7394 & 0.040 & 0.008 & 6.05 & 1.22 & 2.7399 & 0.0004 \\
61355 & 5.0332 & \nodata & \nodata & \nodata & \nodata & \nodata & \nodata \\
210885 & 3.0843 & 5.476 & 0.041 & 484.58 & 7.87 & 3.0828 & \nodata \\
217988 & 1.2214 & -0.055 & 0.062 & -0.39 & 0.44 & 1.2194 & \nodata \\
116352 & 6.7879 & \nodata & \nodata & \nodata & \nodata & \nodata & \nodata \\
\enddata
\tablecomments{
Columns: (1) NIRCam ID; (2) SSP-based systemic redshift from stellar population fitting; (3) H$\alpha$ emission line flux in units of $10^{-17}$ erg s$^{-1}$ cm$^{-2}$; (4) 1$\sigma$ uncertainty on the flux; (5) Rest-frame equivalent width (REW) in \AA; (6) 1$\sigma$ uncertainty on REW; (7) Emission-line redshift derived from the H$\alpha$ fit; (8) 1$\sigma$ uncertainty on line redshift. Only a few rows and columns are shown here. The complete table is available in FITS format on MAST. For detailed description of GELATO fitting outputs, see \citet{hviding_new_2022}. All fits are performed on spectra without slitloss corrections.
}
\end{deluxetable*}
\endgroup

\begingroup
\setlength{\tabcolsep}{8pt}
\begin{deluxetable*}{lcccccccccc}
\tablenum{5}
\tabletypesize{\footnotesize}
\tablewidth{0pt}
\tablecaption{pPXF Line Fit Results for Quiescent Galaxies\label{tab:ppxf_example}}
\tablehead{
\colhead{NIRCam ID} & 
\colhead{$z_{\rm pPXF}$} & 
\colhead{[O\,II] Flux} & 
\colhead{Flux Err} & 
\colhead{H$\beta$ Flux} & 
\colhead{Flux Err} & 
\colhead{[O\,III] 5008} & 
\colhead{Flux Err} & 
\colhead{H$\alpha$ Flux} & 
\colhead{Flux Err}
}
\startdata
204621 & 2.7182 & 0.006 & 0.035 & 0.049 & 0.011 & 0.000 & 0.011 & 0.027 & 0.008 \\
206183 & 1.3170 & \nodata & \nodata & 0.154 & 0.020 & 0.054 & 0.019 & 1.157 & 0.024 \\
209852 & 1.0977 & \nodata & \nodata & 0.000 & 0.020 & 0.025 & 0.018 & 0.159 & 0.017 \\
214075 & 3.6047 & 0.000 & 0.123 & 0.047 & 0.019 & 0.086 & 0.019 & 0.490 & 0.030 \\
217988 & 1.2211 & \nodata & \nodata & 0.007 & 0.028 & 0.077 & 0.025 & 0.013 & 0.031 \\
\enddata
\tablecomments{
pPXF-derived emission line fluxes (in $10^{-17} \rm erg\,s^{-1}cm^{-2}$) for 16 quiescent galaxies in SMILES. Columns are fluxes and associated errors for [O\,II], H$\beta$, [O\,III], and H$\alpha$ (all in erg\,s$^{-1}$\,cm$^{-2}$ units). The complete table is available in FITS format on MAST. All fits are performed on spectra without slitloss corrections.
}
\end{deluxetable*}
\endgroup

\subsection{SED Catalog}

\begin{figure*}
    \includegraphics[width=0.46\linewidth]{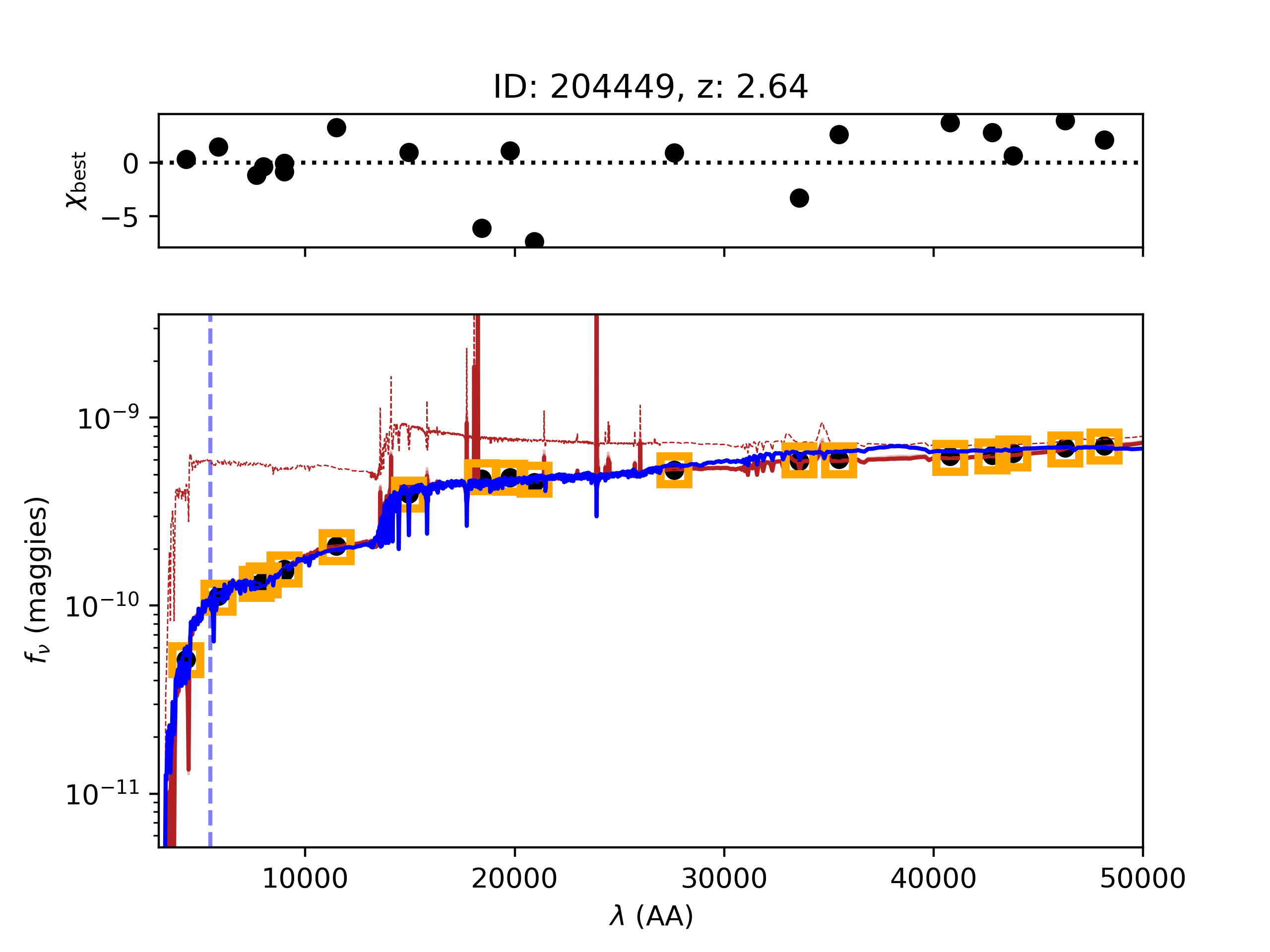}
    \includegraphics[width=0.53\linewidth]{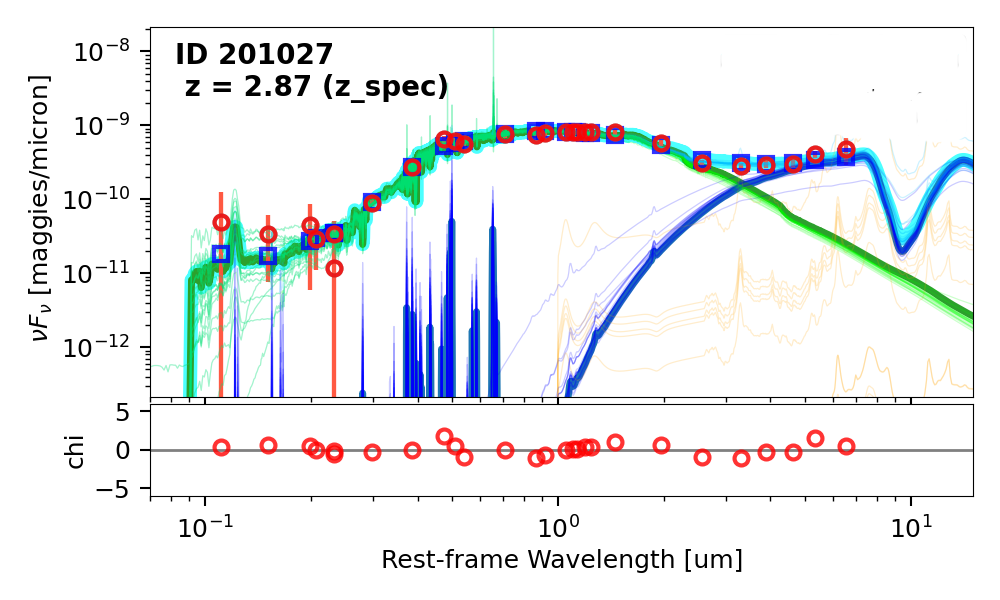}
    \caption{Example SED fits for SMILES galaxies. \textbf{Left:} SED fitting for a non-AGN galaxy (ID: 204449, $z=2.64$) using \texttt{Prospector} on NIRCam photometry only (red line), as adapted from \citet{zhu_higher_2025}. The blue line shows the updated best-fit model including MIRI photometry using the AGN-aware fitting method from \citet{lyu_active_2024}. The dashed red line indicates the intrinsic (dust-free) SED. The vertical dashed blue line marks the location where the UV luminosity is measured ($\lambda_{\rm rest}=1500$ \AA). \textbf{Right:} AGN SED decomposition for a galaxy at $z=2.87$ (JADES ID: 201027), adapted from \citet{rieke_confirming_2025}, showing the combined model (cyan), stellar component (green), AGN torus emission (purple), and stellar-powered infrared excess (orange). Observed photometry is shown with red circles and model photometry with squares.}
    \label{fig:SED}
\end{figure*}

As an ancillary data product, we perform spectral energy distribution (SED) fitting using multi-band photometry to derive the physical properties of galaxies in the SMILES NIRSpec sample. This includes stellar masses, star formation histories, dust attenuation, and metallicities. The SED fitting is performed separately for non-AGN and AGN host galaxies, accounting for their different spectral energy distributions.

For non-AGN galaxies, we use the {\tt Prospector} software \citep{johnson_stellar_2021} combined with the {\tt Parrot} artificial neural network (ANN) emulator\footnote{\url{https://github.com/elijahmathews/MathewsEtAl2023}} \citep{mathews_as_2023} to accelerate the fitting process. The input photometry is taken from the JADES and JEMS surveys \citep{eisenstein_jades_2023, bunker_jades_2024, rieke_jades_2023, williams_jems_2023}, covering a combination of {\it Hubble Space Telescope} (HST) and JWST filters. The photometric dataset includes the HST ACS bands (F435W, F606W, F775W, F814W, F850LP), HST WFC3/IR bands (F105W, F125W, F140W, F160W), and JWST NIRCam bands (F090W, F115W, F150W, F182M, F200W, F210M, F277W, F335M, F356W, F410M, F430M, F444W, F460M, F480M).

The modeling follows the Prospector-$\alpha$ framework \citep{leja_how_2019}, incorporating 14 free parameters that describe a galaxy's stellar mass, gas-phase and stellar metallicity, star formation history (SFH), dust properties, and nebular emission. We adopt a non-parametric SFH model based on the continuity prior \citep{leja_how_2019}, using seven SFH bins to allow for flexible variations in star formation over cosmic time. The fitting assumes a Chabrier initial mass function (IMF) \citep{chabrier_galactic_2003} and models thermal dust emission following \citet{draine_infrared_2007}, with three free parameters controlling PAH mass fraction ($Q_{\rm PAH}$), the minimum radiation field strength for dust emission ($U_{\rm min}$), and the fraction of starlight exposed to radiation fields. The spectroscopic redshift from NIRSpec is used as a prior in the fitting, ensuring accurate physical parameter estimation.

To accelerate the fitting procedure, we use the ANN-based {\tt Parrot} emulator \citep{mathews_as_2023}, which speeds up the SED fitting without introducing significant biases. The emulator produces results consistent with traditional spectral population synthesis models, with deviations of only 25-40\% in stellar mass, metallicity, SFR, and stellar age. The fitting accuracy is validated through comparisons with independent SED models from \citet{ji_jades_2023}, showing deviations of less than 0.15 dex, and with UV-based fits from \citet{lyu_active_2024}, confirming consistency in the rest-frame UV spectral region.

For system with possible AGN evidence, we adopt a modified version of the Prospector code introduced in \citet{lyu_agn_2022,lyu_active_2024} that incorporates semi-empirical AGN models and galaxy dust emission templates. 
These models improve the separation between stellar and AGN components in the observed photometry, leading to more accurate measurements of stellar mass and star formation rates in AGN hosts. An example best-fit SED model is shown in Figure~\ref{fig:SED}. Compared to \citet{lyu_active_2024}, we have enabled nebular emission lines during the fitting process. All other settings follow the same configuration in \citet{lyu_active_2024}, with the default galaxy IR template corresponding to $\log(L_{\mathrm{IR}}/L_\odot) = 11.25$.
We provide the SED properties as a value-added data product through MAST.

\section{SCIENCE DEMONSTRATION} \label{sec:science}
This section demonstrates some initial key scientific analyses enabled by the SMILES NIRSpec dataset. 

\subsection{MIRI-Selected AGN}
The SMILES survey provides a unique opportunity to study obscured AGN by leveraging deep mid-infrared imaging from JWST/MIRI. \citet{lyu_active_2024} identified a population of AGN in the GOODS-S/HUDF field based on their mid-infrared colors and SED properties. However, while mid-infrared selection is highly effective at identifying dusty and obscured AGN, additional spectroscopic diagnostics are necessary to test the efficacy of the photometric selection, i.e. to confirm that the host galaxy properties (e.g., star formation rates) are consistent with the fits and that any AGN emission lines are weak, consistent with the AGN being heavily obscured. The SMILES NIRSpec dataset enables such an investigation by providing rest-frame optical spectra, allowing for a direct comparison using standard emission-line diagnostics such as the BPT diagram (e.g., \citealp{sun_extreme_2025,rieke_confirming_2025}).

\begin{figure*}[!ht]
    \centering
    \includegraphics[width=1.0\linewidth]{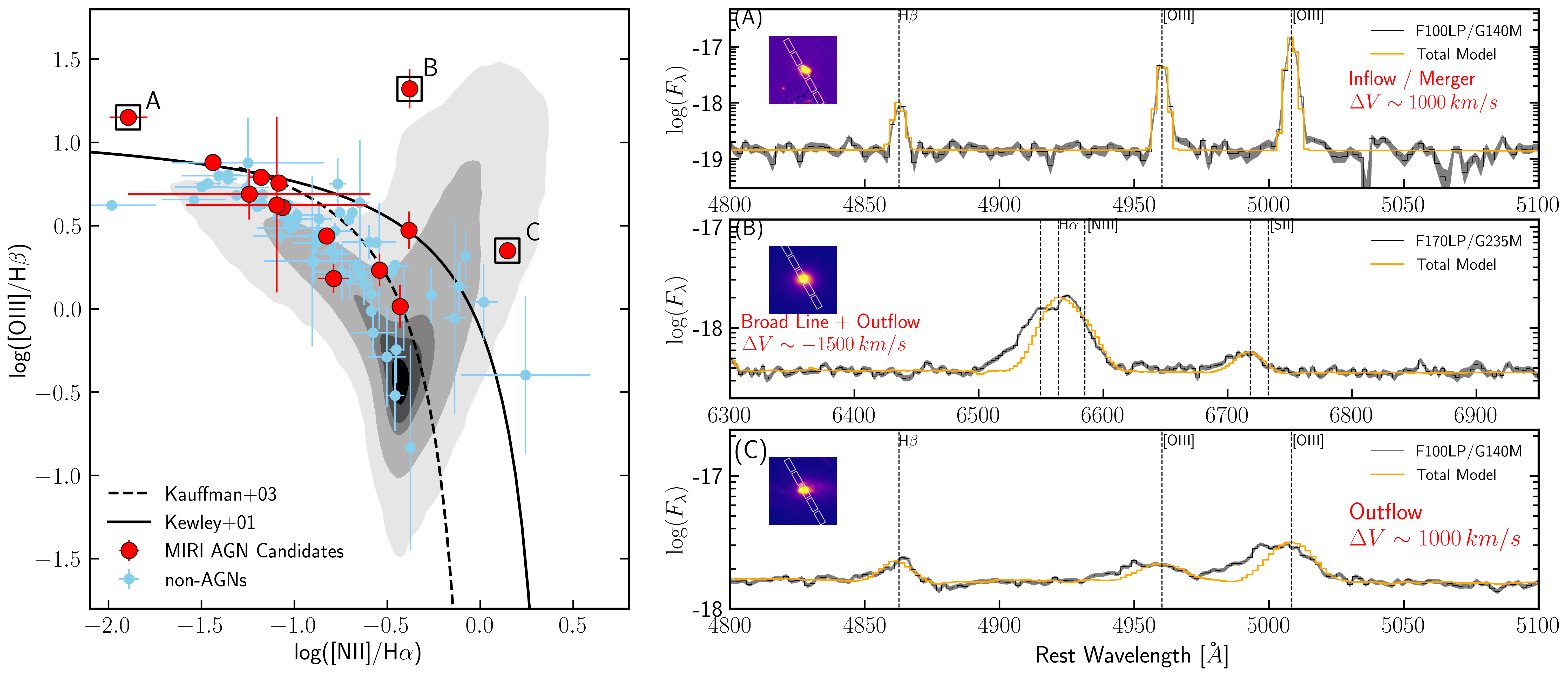}
    \caption{Classification of obscured AGN based on optical region spectra (from \citealp{rieke_confirming_2025}). The SED fit to galaxy (B) shows it to be a lightly obscured Type 1 AGN, consistent with its broad H$\alpha$ line. From the spectra, galaxies (A) and (C) have strong outflows (shown by the similar widths of H $\beta$ and [OIII]). Thus, except for (B), all are consistent with purely star formation in the optical spectrum of the host galaxy, which is consistent with the presence of a heavily obscured AGN with virtually no leakage of UV photons. The hydrogen lines have been used to estimate the star formation rates, which are generally consistent with the values from the SED fits. Both these results support the validity of the SED fitting method to identify obscured AGN.}
    \label{fig:BPT}
\end{figure*}

Figure~\ref{fig:BPT} presents the BPT diagram for a test sample of the MIRI-selected AGN candidates, using key rest-optical emission lines, [O~{\sc iii}] $\lambda5007$/H$\beta$ versus [N~{\sc ii}] $\lambda6584$/H$\alpha$. The red points correspond to the MIRI-selected AGN, while blue points indicate non-AGN galaxies in the SMILES sample. The black solid and dashed curves correspond to the theoretical AGN boundary from \citet{kewley_theoretical_2001} and the empirical classification from \citet{kauffmann_host_2003}, respectively. 

Only the 14 MIRI-selected AGN with adequate signal-to-noise (SNR $>$ 1 in all four lines) are shown. Of these, 11 fall within the star-forming region of the diagram, consistent with the expectation for heavily obscured AGN where ionizing photons are unable to escape and excite the surrounding ISM. This is also consistent with the SED-based AGN classification, which relies on mid-infrared excess rather than optical line ratios. Redshift evolution of the BPT boundary has a minor effect here, as it shifts the demarcation lines slightly upward and to the right, reinforcing the star-forming classification. Three AGN (labeled A: ID 87191, B: ID 209962, and C: ID 196290) are highlighted and shown with their NIRSpec spectra. Galaxy B shows a broad H$\alpha$ profile and UV excess in the SED, consistent with a moderately obscured Type 1 AGN. Galaxies A and C show similarly broadened H$\beta$ and [O~{\sc iii}] lines, indicative of gas flows, likely outflows or mergers. Both also exhibit strong mid-infrared excesses well above the stellar photospheric contribution, consistent with deeply embedded AGN activity.

\subsection{Multi-Phase Outflows from Cosmic Noon Galaxies}

\begin{figure*}[!ht]
    \centering
    \includegraphics[width=0.63\linewidth]{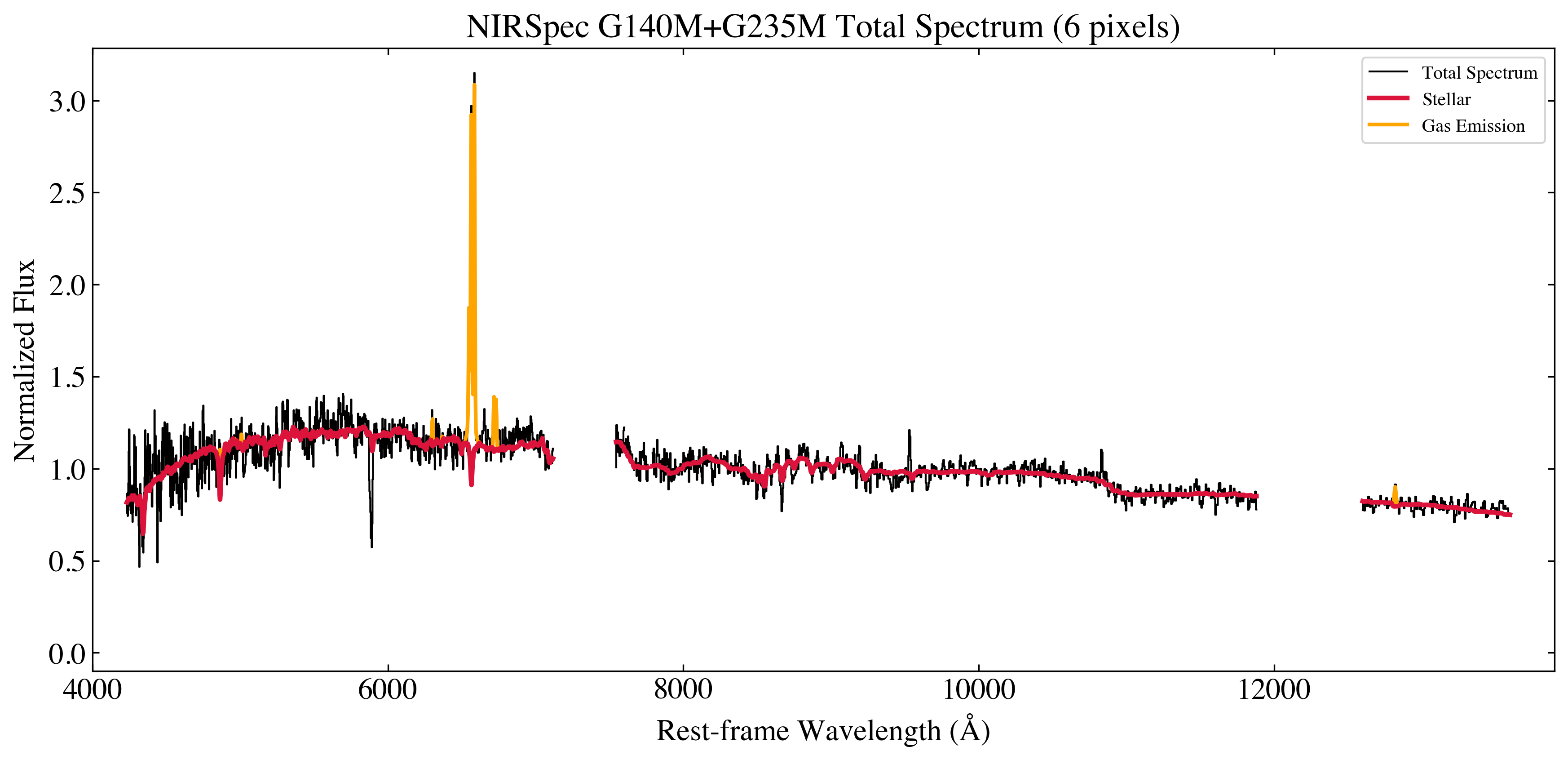}
    \includegraphics[width=0.33\linewidth]{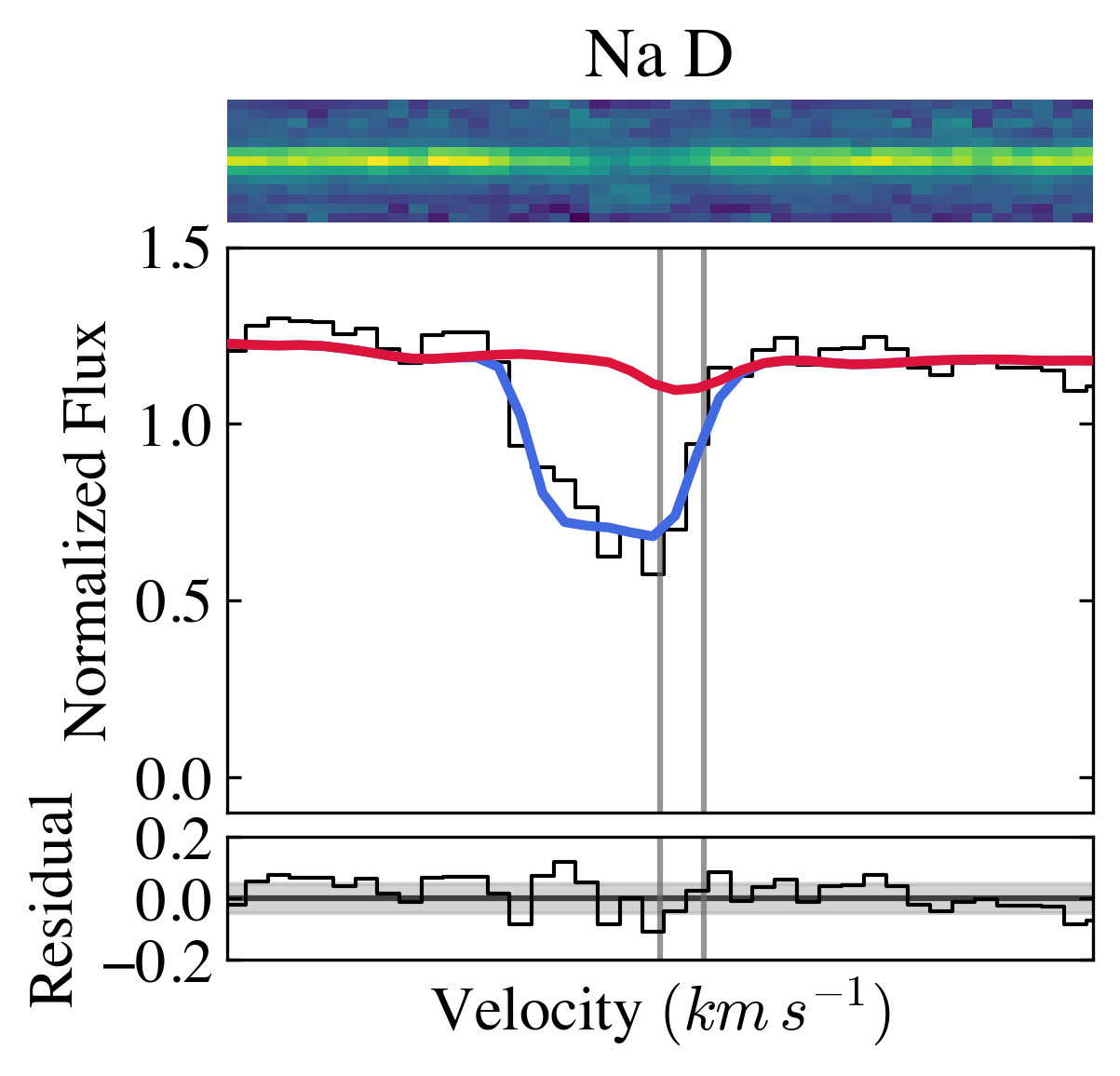}
    \caption{Neutral outflow traced by Na\,{\sc D} in galaxy ID-206183 at $z=1.3171$ \citep[from][]{sun_extreme_2025}. 
    \textbf{Left:} NIRSpec G140M+G235M spectrum showing the rest-optical features, including Balmer lines, and a clear absorption feature at the Na\,{\sc D} doublet. Best fits for the stellar continuum and gas emission features \added{from {\tt pPXF}} are also shown in red and yellow lines, respectively.
    \textbf{Right:} Zoom-in on the Na\,{\sc D} region. The top panel shows the 2D spectrum; the middle panel shows the normalized 1D spectrum with the best-fit absorption profile (red) and decomposed outflow component (blue); the bottom panel displays the residual. A blueshifted absorption trough indicates the presence of a strong neutral outflow component.}
    \label{fig:outflows}
\end{figure*}

Galactic outflows play a crucial role in regulating star formation and metal enrichment in galaxies across cosmic time. The SMILES NIRSpec dataset provides an opportunity to systematically study ionized and neutral outflows at $z\sim1$-3, tracing multiple gas phases and their driving mechanisms. 

\textbf{Ionized outflows} in the SMILES sample are identified through a two-step process. First, outflow candidates are selected based on extended nebular emission in JWST medium-band imaging \citep{zhu_systematic_2025}, where galaxies exhibiting significantly larger extents in emission-line-sensitive filters compared to adjacent continuum bands are flagged as potential outflow hosts. A subset of these candidates has available SMILES NIRSpec spectra, allowing us to confirm the presence of ionized outflows through velocity shifts and asymmetric broadening of key rest-frame optical emission lines. The most prominent tracers of these outflows are the [O~{\sc iii}] $\lambda5007$ and H$\alpha$ lines, which frequently exhibit significant blueshifted components indicative of fast-moving ionized gas. These velocity offsets provide a direct measure of the gas kinematics, with outflow candidates showing a median blueshift of $\Delta v=-95\,\rm km\,s^{-1}$, compared to $\Delta v=-5\,\rm km\,s^{-1}$ in a redshift-matched control sample of galaxies without extended emission-line features. The presence of high-velocity ionized gas suggests that these outflows are driven by feedback processes, either from intense star formation or AGN. Furthermore, AGN-hosting candidates in the SMILES sample tend to show more anisotropic extended emission compared to star-forming galaxies, suggesting that AGN-driven outflows may have a preferred directional structure. These results highlight the power of JWST in identifying and characterizing ionized outflows, with medium-band imaging serving as an efficient pre-selection method and NIRSpec spectroscopy providing robust velocity diagnostics to confirm and study the kinematics of ionized gas in galaxies at cosmic noon. Examples of spectra with AGN-driven outflows are also shown in Figure \ref{fig:BPT}.

\textbf{Neutral outflows}, traced by the Na~D $\lambda\lambda$5890, 5896 absorption doublet, provide a critical window into the ejection of cold gas from galaxies. With much higher sensitivity and NIR wavelength coverage of JWST/NIRSpec, Na D-traced outflows can now be detected and systematically studied at cosmic noon \citep[e.g.,][]{davies_jwst_2024}.
\added{We identify 14 galaxies with significant Na~D ISM absorption, including 4 outflows, 7 with systemic interstellar medium (ISM) absorption, 2 with inflowing gas, and 1 showing an emission/P-Cygni profile (Y. Sun et al. in prep). Combined with the Blue Jay Na D sample \citep{davies_jwst_2024} and JADES observations, this dataset represents the largest systematic study of Na~D-traced neutral outflows at $z\sim2$. }
In the SMILES dataset, we discovered one of the strongest Na~D outflows observed beyond $z\sim1$ in the quenching galaxy JADES-GS-206183 (Figure \ref{fig:outflows}). Its SMILES NIRSpec spectrum reveals a significantly blueshifted and deep Na~D absorption feature, indicating an outflow velocity of $v_{\rm out}=828^{+79}_{-49}\,\mathrm{km\,s^{-1}}$ and a substantially high mass outflow rate of $\log(\dot{M}_{\rm out})=2.40^{+0.11}_{-0.16}\,\mathrm{M_{\odot}\,yr^{-1}}$. This extreme outflow is comparable to those found in local galaxies undergoing intense starbursts or hosting luminous AGN. However, SED modeling of JADES-GS-206183 based on multi-band photometry from HST/ACS to JWST/NIRCam suggests that it is now quenched, with a low ongoing star formation rate confirmed by its Paschen $\alpha$ (Pa$\alpha$) emission in the FRESCO NIRCam grism spectrum ($10.78\pm 0.55\,\mathrm{M_{\odot}\,yr^{-1}}$). Furthermore, optical line ratio diagnostics indicate weak AGN activity, making it unlikely that the current outflow is driven by ongoing star formation or nuclear activity. Instead, this system likely hosts a long-lasting fossil outflow, powered by past AGN feedback that has now faded \citep{sun_extreme_2025}. 

These results highlight the strong capability of JWST/NIRSpec for studying multi-phase outflows in individual galaxies through rest-frame optical emission or absorption lines beyond z$\sim$1, as well as the importance of multi-phase outflow diagnostics in understanding galaxy quenching and feedback mechanisms in the early universe and their cosmic evolution.

\subsection{Ionizing Properties of Galaxies}

\begin{figure}
    \centering
    \includegraphics[width=1.0\linewidth]{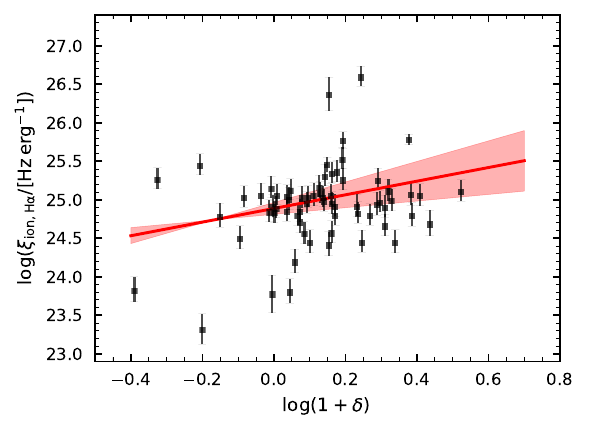}
    \caption{Trend between ionizing photon production efficiency ($\xi_{\rm ion}$) and galaxy overdensity, measured from SMILES NIRSpec medium-resolution spectra combined with JADES multi-band photometric overdensity measurements (adapted from \citealt{zhu_higher_2025}).}
    \label{fig:xion}
\end{figure}

Understanding where and how galaxies efficiently produce ionizing photons is key to constraining the topology of reionization \citep[e.g.,][]{becker_evidence_2018,christenson_relationship_2023,cain_chasing_2025}. \citet{zhu_higher_2025} used the SMILES NIRSpec spectra to investigate the relationship between ionizing photon production efficiency, $\xi_{\rm ion}$ \citep{robertson_cosmic_2015}, and galaxy overdensity, $\log(1+\delta)$, by combining spectroscopic measurements with overdensity maps derived from the JADES multi-band photometric catalog.

\citet{zhu_higher_2025} analyzed a sub-sample of 79 galaxies at $1.0 < z < 5.2$, of which 67 have H$\alpha$ coverage ($1.0 < z < 3.1$), enabling accurate determination of $\xi_{\rm ion}$ from H$\alpha$ luminosities. As shown in Figure~\ref{fig:xion}, $\xi_{\rm ion}$ exhibits a positive trend with galaxy overdensity, and this represents the first such measurement. On average, galaxies in overdense regions ($\log(1+\delta) > 0.1$) produce 2.43 times more ionizing photons than galaxies in lower-density environments. This trend is independent of redshift evolution over the sampled range (see also \citealp{li_reionization_2025}). \citet{zhu_higher_2025} also confirmed strong correlations between $\xi_{\rm ion}$ and the rest-frame equivalent widths of [O~{\sc iii}] and H$\alpha$, consistent with previous results \citep[e.g.,][]{tang_mmtmmirs_2019,pahl_spectroscopic_2024}. No significant dependence was found between $\xi_{\rm ion}$ and the O$_{32}$ ratio or $M_{\rm UV}$.

This analysis highlights the unique synergy between JWST/NIRSpec MSA spectroscopy and deep NIRCam photometric redshift catalogs, enabling environment-based studies of ionizing properties for large galaxy samples. The observed $\xi_{\rm ion}$-overdensity trend suggests that galaxies in high-density environments are systematically more efficient at producing ionizing photons. If this behavior persists at higher redshifts ($z \sim 6$), it may help explain enhanced Ly$\alpha$ transmission observed near galaxy overdensities during reionization \citep[e.g.,][]{jin_spectroscopic_2024}. Future large-area JWST surveys will be helpful for extending this analysis to larger cosmic volumes and for further constraining the role of environment in shaping galaxy ionizing output.

\subsection{Quiescent Galaxies}

The SMILES sample includes 16 massive QG candidates spanning a redshift range of $z = 1.10$ to $3.60$. These galaxies are selected based on their rest-optical continuum morphologies and spectral energy distributions, and confirmed through NIRSpec observations. The pPXF fits (Table \ref{tab:ppxf_example}) provide stellar redshifts and strong constraints on weak nebular emission. Remarkably, 15 out of the 16 QGs show detectable H$\alpha$ emission at the $>1\sigma$ level, enabling investigations of low-level residual star formation or AGN-related ionized gas in otherwise passive systems. Combined with measurements of [O\,III], H$\beta$, and [O\,II] where available, this subset offers a unique opportunity to explore the diversity of ionization conditions and gas retention in quiescent galaxies during the epoch when massive galaxies are first shutting down star formation. A detailed analysis will be presented in Zhiyuan Ji et al. (in prep).

\section{Summary} \label{sec:summary}

In this paper, we present the second data release of the SMILES survey, focusing on JWST/NIRSpec medium-resolution spectroscopy of galaxies spanning $0 \lesssim z \lesssim 7.5$. This dataset provides one of the largest medium-resolution spectroscopic samples of galaxies at cosmic noon and beyond, complementing the deep mid-infrared imaging from MIRI.

We describe the sample selection and observational strategy, designed to target a diverse population of star-forming galaxies, quiescent galaxies, and AGN. The observations employ the G140M/F100LP and G235M/F170LP gratings, covering a wavelength range of $0.97\,\mu{\rm m} < \lambda < 3.07\,\mu{\rm m}$ with a spectral resolution sufficient for detailed emission-line diagnostics.

We outline the data reduction process, which includes detector-level corrections, spectroscopic calibration, and optimal spectral extraction. Custom processing steps are applied to mitigate instrumental effects such as $1/f$ noise and slit loss, ensuring high-quality spectra. The final data products include calibrated 2D and 1D spectra, a spectroscopic redshift catalog, an emission-line catalog, and SED fits based on multi-band photometry.

The spectroscopic dataset enables a wide range of scientific investigations. We highlight opportunities for studying obscured AGN, galactic outflows, and the evolution of gas-phase metallicity in high-redshift galaxies. The emission-line catalog allows for robust measurements of ionized gas conditions, star formation rates, and AGN activity, while the redshift catalog provides a well-validated reference for studies of cosmic structure formation.

The SMILES NIRSpec dataset represents a significant step forward in the study of galaxy evolution, leveraging JWST’s unprecedented sensitivity and spectral resolution. Future works can focus on further characterization of galaxy properties, comparison with theoretical models, and integration with complementary datasets from ALMA, HST, and deep-field ground-based surveys. The full dataset and catalogs are available through STScI/MAST.

\begin{acknowledgments}
\added{We thank the anonymous reviewer for their helpful feedback.}
Y.Z., Y.S., J.M.H., Z.J., M.J.R., and C.N.A.W. acknowledge support from the NIRCam Science Team contract to the University of Arizona, NAS5-02105. G.H.R., S.A., J.L., and J.M. acknowledge support from the JWST Mid-Infrared Instrument (MIRI) Science Team Lead, grant 80NSSC18K0555, from NASA Goddard Space Flight Center to the University of Arizona. I.S. acknowledges fundings from the European Research Council (ERC) DistantDust (Grant No.101117541) and the Atracc\'{i}on de Talento Grant No.2022-T1/TIC-20472 of the Comunidad de Madrid, Spain

This work is based on observations made with the NASA/ESA/CSA James Webb Space Telescope. The data were obtained from the Mikulski Archive for Space Telescopes at the Space Telescope Science Institute, which is operated by the Association of Universities for Research in Astronomy, Inc., under NASA contract NAS 5-03127 for JWST. These observations are associated with programs \#1207. The specific observations analyzed can be accessed via \url{https://archive.stsci.edu/hlsp/smiles} under DOI \url{https://doi.org/10.17909/et3f-zd57}.

We respectfully acknowledge the University of Arizona is on the land and territories of Indigenous peoples. Today, Arizona is home to 22 federally recognized tribes, with Tucson being home to the O’odham and the Yaqui. The university strives to build sustainable relationships with sovereign Native Nations and Indigenous communities through education offerings, partnerships, and community service.

This manuscript benefited from grammar checking and proofreading using ChatGPT \citep{openai_chatgpt_2023}.
\end{acknowledgments}


\vspace{5mm}
\facilities{JWST, MAST}

\software{
{\tt JWST Calibration Pipeline} \citep{bushouse_jwst_2022}
}



\bibliographystyle{aasjournal}

\end{document}